\documentclass[preprint,superscriptaddress,aps,eqsecnum]{revtex4}
\usepackage{amsfonts}
\usepackage{graphics}
\usepackage{graphicx}
\newcommand{\bea}{\begin{eqnarray}}
\newcommand{\eea}{\end{eqnarray}}
\begin{document}
\title{On the dual equivalence between self-dual and Maxwell-Chern-Simons models with Lorentz symmetry breaking}
\author{C. Furtado, J. R. Nascimento, A. Yu. Petrov}
\affiliation{Departamento de F\'{\i}sica, Universidade Federal da Para\'{\i}ba\\
 Caixa Postal 5008, 58051-970, Jo\~ao Pessoa, Para\'{\i}ba, Brazil}
\email{furtado, jroberto, petrov@fisica.ufpb.br}
\author{M. A. Anacleto}
\affiliation{Universidade Federal do Piau\'{\i}, Campus Ministro Reis Velloso, 64202-020, Parna\'{\i}ba, Piau\'{\i}, Brazil}
\email{anacleto@ufpi.edu.br}
\begin{abstract}
In this paper, we use gauge embedding procedure and master action approach to establish the equivalence between the self-dual and Maxwell-Chern-Simons models with Lorentz symmetry breaking. As a result, new kinds of Lorentz-breaking terms arise.
\end{abstract}
\maketitle
\pretolerance10000
\section{Introduction}
The hypothesis about the possibility of the Lorentz symmetry breaking is an important ingredient of the modern quantum field theory. Being initially inspired by the study of the cosmic rays \cite{GZK}, further it received more solid motivations from cosmological studies \cite {Mag} and development of the noncommutative field theory \cite{CarKos}. The Lorentz symmetry breaking was shown to have a lot of important physical conclusions, such as possibility of arising of new classes of terms in Lagrangians \cite{Kostel}, modification of the dispersion relations, birefringence of light in a vacuum and rotation of plane of polarization of light in a vacuum (some papers devoted to these results are given in \cite{Ja}) and many other consequences.

Most of these implications of the Lorentz symmetry breaking were obtained in four-dimensional space-time where the electrodynamics with Jackiw term (see f.e. \cite{JK}) plays the role of the standard Lorentz-breaking theory whose different aspects were studied in \cite{list} (nevertheless, the Lorentz breaking was studied also for other four-dimensional theories, such as, for example, linearized and non-linearized gravity \cite{gra}). At the same time, there is much less results for the Lorentz-breaking theories in other space-time dimensions, the only results are the study of compactification of the five-dimensional Lorentz-breaking theories \cite{Obousy}, the study of two-dimensional Lorentz-breaking model for the scalar fields \cite{our2d} and investigation of some phenomenological implications of the three-dimensional "mixed" scalar-vector quadratic term \cite{JT} which was earlier obtained via the dimensional reduction of the Jackiw term \cite{JTred0,JTred}. So, the natural problem is the investigation of more aspects of the lower-dimensional, especially three-dimensional, Lorentz-breaking field theories.

One of the important phenomena taking place in three-dimensional field theories is the duality between self-dual and Maxwell-Chern-Simons theories \cite{indual}. Different aspects of the duality (including the supersymmetric case) were studied in a number of papers \cite{dual,mal,embed} (it should be noted that the duality of the four-dimensional theories, which must involve Lorentz symmetry breaking was also studied, see \cite{4dual}). Thus, it seems to be that the very interesting problem is the generalization of a duality for the Lorentz-breaking theories. This problem is the main object of study in this paper. Here we construct the Lorentz-breaking analog of the self-dual model, carry out the gauge embedding algorithm \cite{embed} and develop the master action approach \cite{mal} and obtain a new Lorentz-breaking theory whose important ingredient is the "mixed" scalar-vector quadratic term.

\section{Dual embedding for free Lorentz-breaking self-dual model}
Let us introduce the following Lagrangian for the three-dimensional self-dual model with Lorentz symmetry breaking:
\begin{eqnarray}
\label{acao}
{\cal L}&=&\frac{m}{2} \epsilon^{\mu\nu\rho}f_{\mu}\partial_{\nu}f_{\rho}-\frac{m^2}{2}f_{\mu}f^{\mu} 
+\frac{1}{2}\partial_{\mu}\phi\partial^{\mu}\phi+ 2m\phi v^{\mu}f_{\mu}.
\end{eqnarray}
This Lagrangian is quite similar to that one used in the first paper \cite{embed} treating nonsupersymmetric theory. However, it has an essential difference, that is, the Lorentz symmetry breaking is implemented via the term $2m\phi v^{\mu}f_{\mu}$, where the constant 3-vectors $v_{\mu}$ introduced the preferred frame in the space-time.
  
The Lagrangian equations of motion for this model look like
\begin{equation}
\label{eqmov}
m\epsilon_{\mu\nu\rho}\partial^{\nu}f^{\rho}-m^2f_{\mu}+2m\phi v_{\mu}=0
\end{equation}

Now, we turn to study of duality between the self-dual and Maxwell-Chern-Simons (MCS) models with Lorentz symmetry breaking.
To establish the equivalence of these theories, we use the iterative gauge embedding procedure~\cite{embed}. This is done by extension of the original Lagrangian by the additive terms depending of the Euler vectors $K_{\mu}$, i.e. the left-hand sides of the equations of motion:
\begin{equation}
{\cal L}\rightarrow {\cal L} + F(K_{\mu})
\end{equation}
where the original Lagrangian is given by (\ref{acao}), and $F(K_{\mu})$ is such that $F(0)=0$.   

The variation of the Lagrangian (\ref{acao}) with respect to $f_{\mu}$ leads to the Euler vectors $K_{\mu}$:
\begin{equation}
K^{\mu}=-m^2f^{\mu}+m\epsilon^{\mu\nu\rho}\partial_{\nu}f_{\rho}+2m\phi v^{\mu},
\end{equation}
with the equations of motion are given by the condition $K_{\mu}=0$.

Let us follow the gauge embedding approach similarly to \cite{embed}. Since our aim is to obtain the gauge invariant theory, let us suggest that the desired gauge transformation for the vector field $f_{\mu}$ be $\delta f_{\mu}=\partial_{\mu}\epsilon$, with $\epsilon$ is a parameter of gauge transformations. Thus, the variation of the Lagrangian under these transformations is $\delta {\cal L}=K^{\mu}\partial_{\mu}\epsilon$. Then, we introduce the first-order iterated Lagrangian
\begin{equation}
{\cal L}^{(1)}={\cal L}-\Lambda^{\mu}K_{\mu},
\end{equation}
where $\Lambda^{\mu}$ is a Lagrange multiplier. We suppose the gauge transformation for the $\Lambda^{\mu}$ to be $\delta\Lambda_{\mu}=\partial_{\mu}\epsilon$ which we choose to cancel the variation of ${\cal L}$ (cf. \cite{embed}).
Thus, the variation of ${\cal L}^{(1)}$ under the gauge transformations is $\delta{\cal L}^{(1)}=-\Lambda^{\mu}\delta K_{\mu}$; since $\delta K_{\mu}=-m^2\partial_{\mu}\epsilon$, we find 
$\delta{\cal L}^{(1)}=m^2\Lambda^{\mu}\partial_{\mu}\epsilon=\frac{m^2}{2}\delta(\Lambda^{\mu}\Lambda_{\mu})$.
To cancel this term, we add to the Lagrangian the term $-\frac{m^2}{2}\delta(\Lambda^{\mu}\Lambda_{\mu})$, thus obtaining the second-order iterated, gauge invariant Lagrangian
\begin{eqnarray}
{\cal L}^{(2)}={\cal L}-\Lambda^{\mu}K_{\mu}-\frac{m^2}{2}\Lambda^{\mu}\Lambda_{\mu},
\end{eqnarray}
which after elimination of the auxiliary field $\Lambda_{\mu}$ via its equations of motion (which look like $K_{\mu}=-m^2\Lambda_{\mu}$) gets the form
\begin{eqnarray}
\label{leff}
{\cal L}_{eff}&=&{\cal L}+\frac{1}{2m^2}K^{\mu}K_{\mu}
\nonumber\\
&=&\frac{1}{2}F_{\mu}F^{\mu}-\frac{m}{4}\epsilon^{\mu\nu\rho}A_{\mu}F_{\nu\rho}
+\frac{1}{2}\partial_{\mu}\phi\partial^{\mu}\phi
+\phi\epsilon^{\mu\nu\rho} v_{\mu}F_{\nu\rho}+2\phi^2v_{\mu}v^{\mu}.
\end{eqnarray}
where we have renamed, $f_{\mu}\rightarrow A_{\mu}$, to reflect the invariant character of the theory. 
Here
\begin{equation}
F^{\mu}\equiv\frac{1}{2}\epsilon^{\mu\nu\alpha}F_{\nu\alpha},
\end{equation}
is the dual of the tensor $F_{\nu\alpha}$. Thus, we succeeded to construct the dual projection of the Lorentz-breaking self-dual model.

To estabilsh the duality, it remains to compare the equations of motion for the matter sector of both models, that is, self-dual one (\ref{acao}) and the Maxwell-Chern-Simons one (\ref{leff}). The equations of motion to the scalar field, $\phi$, of the self-dual model reads,
\begin{equation}
\partial_{\mu}\partial^{\mu}\phi=2mv_{\mu}f^{\mu}.
\end{equation}
From the MCS model we find the equations for the field $\phi$,
\begin{equation}
\partial_{\mu}\partial^{\mu}\phi=2mv_{\mu}\left[\frac{F^{\mu}}{m}+\frac{2\phi}{m}v^{\mu}\right].
\end{equation}
Comparing the right-hand sides of these equations, we finally obtain that the correct map from self-dual model to the Maxwell-Chern-Simons one which is given by the following relation between the vector fields of two models: 
\begin{equation}
\label{dmap}
f^{\mu}\rightarrow \frac{F^{\mu}}{m}+\frac{2\phi}{m}v^{\mu}.
\end{equation}
Thus, the constructing of the dual mapping of the Lorentz-breaking self-dual model and the Lorentz-breaking Maxwell-Chern-Simons model is complete.


It is also interesting to study dispersion relations of the Maxwell-Chern-Simons theory we obtained and the self-dual theory. First we turn to the Maxwell-Chern-Simons theory we obtained. We note that the theory studied in \cite{JTred0} involves a massless scalar field, thus, our result will differ from that one from \cite{JTred0} reproducing last one in the case of the light-like $v^{\mu}$.

Using the coefficients (\ref{pmcs}) and (\ref{psd}) of the expansion of the propagators (see Appendix), we find that the dispersion relations corresponding to the propagator of the Maxwell-Chern-Simons theory are: first, common Lorentz-invariant massless one, $E^2=\vec{p}^2$, second, common Lorentz-invariant massive one $E^2=\vec{p}^2+m^2$, third, for the space-like or light-like $v^{\mu}$, also the Lorentz-invariant one $E^2=\vec{p}^2+4v^2$, fourth, the Lorentz-violating one, produced by the condition ${\cal R}=0$: $(E^2-\vec{p}^2-m^2)(E^2-\vec{p}^2-M^2)+v^2(E^2-\vec{p}^2)+(\vec{v}\cdot\vec{p}-v_0E)^2=0$, with $M^2=4v^2$.

For the self-dual theory, the corresponding dispersion relations are again first, common Lorentz-invariant massless one, $E^2=\vec{p}^2$, second, common Lorentz-invariant massive one $E^2=\vec{p}^2+m^2$. However, third dispersion relation, unlike of the Maxwell-Chern-Simons case, is also the Lorentz-invariant one $(E^2-\vec{p}^2)^2-(E^2-\vec{p}^2)m^2+4m^2v^2=0$. Thus, one can conclude that the physical states in the self-dual theory are Lorentz-invariant, so, dual embedding of the self-dual theory modifies the dispersion relations in a nontrivially Lorentz-breaking way whereas in the case of the self-dual theory the dispersion relations are Lorentz invariant, the only impacts of the Lorentz-breaking vector $v_{\mu}$ is in the numerator of the propagator and in the modification of the mass. From a formal viewpoint this is related by the fact that in the self-dual theory the $v_{\mu}$ enters the denominator only in the form of an invariant square $v^2$ whereas in the case of the MCS theory -- within the object $T^{\mu}T_{\mu}$ which evidently introduces the preferential directions. At the same time, it should be noted that difference of the mass spectra of the dual theories is not an unusual fact since only the physical sectors of spectra of the dual theories must coincide. 

Indeed, the propagators of the both theories, being both of the form $\Delta$ (\ref{del1}), but with different $M_{\mu\nu}$ and $T_{\mu}$, are the Hermitian operators which can be simultaneously transformed to the diagonal form. Afterwards, the dispersion relations do not change, persisting to be of the same form as above. Imposing an appropriate gauge for the Maxwell-Chern-Simons theory and solving constraints for the self-dual theory, we can eliminate the irrelevant degrees of freedom corresponding to the nonphysical sector, thus remaining with the only physical particles whose dispersion relations in both theories look like $E^2=\vec{p}^2$ and $E^2=\vec{p}^2+m^2$, for two physical degrees of freedom. The detailed study of the unitarity and causality aspects of the Lorentz-breaking Maxwell-Chern-Simons theory (\ref{leff}) within which the nonphysical sector is shown to decouple, was carried out in \cite{JTred0}  for the case of the $M^2=0$ which corresponds to the case of the light-like $v^{\mu}$, and can be straightforwardly generalized for the case $M^2\neq 0$ (see also \cite{Baeta} for general issues related to the problems of unitarity and causality in Lorentz-breaking theories).

\section{Dual embedding for the Lorentz-breaking self-dual theory coupled to the spinor matter}

Let us extend the self-dual Lorentz-breaking model via coupling of the vector field to the extra spinor matter. So, we introduce the current $j^{\mu}=\bar{\psi}\gamma^{\mu}\psi$, hence the Lagrangian be
\begin{eqnarray}
\label{ac1}
{\cal L}&=&\frac{m}{2} \epsilon^{\mu\nu\rho}f_{\mu}\partial_{\nu}f_{\rho}-\frac{m^2}{2}f_{\mu}f^{\mu} 
+\frac{1}{2}\partial_{\mu}\phi\partial^{\mu}\phi+ 2m\phi v^{\mu}f_{\mu}+f^{\mu}j_{\mu}.
\end{eqnarray}
The corresponding Euler vector for the vector field is
\begin{equation}
K_{\mu}=m\epsilon_{\mu\nu\rho}\partial^{\nu}f^{\rho}-m^2f_{\mu}+2m\phi v_{\mu}+j_{\mu}.
\end{equation}
We can proceed with the gauge embedding algorithm as in the previous section. As a result, we arrive at the following second-order iterated Lagrangian:
\begin{eqnarray}
\label{leff1}
{\cal L}_{eff}&=&{\cal L}+\frac{1}{2m^2}K^{\mu}K_{\mu}
\nonumber\\
&=&\frac{1}{2}F_{\mu}F^{\mu}-\frac{m}{4}\epsilon^{\mu\nu\rho}A_{\mu}F_{\nu\rho}
+\frac{1}{2}\partial_{\mu}\phi\partial^{\mu}\phi
+\phi\epsilon^{\mu\nu\rho} v_{\mu}F_{\nu\rho}+\nonumber\\&+&
\frac{1}{2m^2}j^{\mu}j_{\mu}+\frac{1}{m}j^{\mu}F_{\mu}+\frac{2}{m}\phi v^{\mu}j_{\mu}+2\phi^2v_{\mu}v^{\mu}.
\end{eqnarray}
We find that, due to coupling of the vector field to the spinor field, we find, first, a Thirring-like current-current interaction, second, a "magnetic" coupling of the matter to the vector field, third, a new, Lorentz-breaking coupling of the spinor matter to the scalar field.

In this case, the analog of the dual mapping (\ref{dmap}) looks like
\begin{equation}
\label{dmap1}
f^{\mu}\rightarrow \frac{F^{\mu}}{m}+\frac{2\phi}{m}v^{\mu}+\frac{1}{m^2}j^{\mu},
\end{equation}
thus, the dual projection of the self-dual field depends on electromagnetic field, spinor matter and Lorentz-breaking vector.
We note that for the spinor matter current $j^{\mu}$, generalization for the noncommutative case is straightforward.

\section{Duality of two models within the master action approach}

Let us show the duality of the self-dual Lorentz-breaking model coupled to the matter (\ref{ac1}) and of the Maxwell-Chern-Simons Lorentz-breaking model coupled to the matter (\ref{leff1}) in a way similar to \cite{mal}. First of all, we find that there is a dual identification $f^{\mu}\to \frac{1}{m}F^{\mu}$, as in \cite{dual}. Second, to confirm the duality we can introduce a master Lagrangian
\bea
{\cal L}_{master}&=&-\frac{m^2}{2}f^{\mu}f_{\mu}+mf^{\mu}F_{\mu}-\frac{m}{2}F^{\mu}A_{\mu}+\frac{1}{2}\partial_{\mu}\phi\partial^{\mu}\phi+f_{\mu}(2m\phi v^{\mu}+j^{\mu})-\nonumber\\&-&\frac{1}{2\xi}(\partial_{\mu}A^{\mu})^2.
\eea
If one integrates over the fields $f^{\mu}$, the result be
\bea
{\cal L}_{MCS}^{eff}&=&\frac{1}{2}F^{\mu}F_{\mu}-\frac{m}{4}\epsilon^{\mu\nu\rho}A_{\mu}F_{\nu\rho}-
\frac{1}{2\xi}(\partial_{\mu}A^{\mu})^2
+\frac{1}{m}F^{\mu}(j_{\mu}+2m\phi v_{\mu})+\nonumber\\&+&\frac{1}{2m^2}(j^{\mu}+2m\phi v^{\mu})(j_{\mu}+2m\phi v_{\mu})+\frac{1}{2}\partial_{\mu}\phi\partial^{\mu}\phi,
\eea
which reproduces the Lagrangian (\ref{leff1}).

At the same time, if one integrates over the fields $A_{\mu}$, one arrives at
\bea
{\cal L}_{SD}^{eff}&=&-\frac{m^2}{2}f^{\mu}f_{\mu}+\frac{m}{2}\epsilon^{\mu\nu\rho}f_{\mu}\partial_{\nu}f_{\rho}+f_{\mu}
(2m\phi v^{\mu}+j^{\mu})+\frac{1}{2}\partial_{\mu}\phi\partial^{\mu}\phi,
\eea
which reproduces the Lagrangian (\ref{ac1}). Thus, we confirmed the duality of these theories. It is clear that after the integration over the remaining vector fields they imply in the same generating functionals, that is
\bea
Z[j,\phi]&=&\exp\left(-\frac{i}{2}(2m\phi v^{\mu}+j^{\mu})\frac{1}{\Box-m^2}[\eta_{\mu\nu}-\frac{\partial_{\mu}\partial_{\nu}}{m^2}-\frac{1}{m}\epsilon_{\mu\nu\lambda}\partial^{\lambda}](2m\phi v^{\nu}+j^{\nu})+\right.\nonumber\\&+&\left.\frac{i}{2}\partial_{\mu}\phi\partial^{\mu}\phi\right).
\eea
Thus, the proof of equivalence is completed. Indeed, we have shown that the Lagrangians (\ref{ac1}) and (\ref{leff1}) imply in the same quantum dynamics.

\section{Summary}

Let us discuss our results. We succeeded, via the gauge embedding method, to construct a new Lorentz-breaking theory described by the Lagrangian (\ref{leff}), with further this duality was confirmed via master action approach. First of all, we find that it involves not only the massive term for the vector field, which is the well-known Chern--Simons term (the similar situation takes place in the Lorentz-invariant case \cite{embed}), but also the
massive term for the scalar field, that is, the last term in eq. (\ref{leff}), which is fundamental to maintain the contents of the scalar sectors unchanged, thus, the gauge embedding generates the mass both for the vector field and for the matter field.  Second, it includes the  desired "mixed" scalar-vector term \cite{JT} $\phi\epsilon^{\mu\nu\rho} v_{\mu}F_{\nu\rho}$ which earlier was obtained via dimensional reduction \cite{JTred0}.

This "mixed" term possesses the "restricted" gauge symmetry, that is, only the vector field is transformed under the gauge transformations whereas the matter field remains unchanged. However, this is very natural since the gauge embedding algorithm requires that the matter field should be unchanged \cite{embed}. Indeed, even in the Lorentz-invariant theories \cite{embed} the action obtained after the gauge embedding procedure also possessed only restricted gauge symmetry, thus, the "restricted" gauge invariance of the theory obtained in this case is very natural.

We have studied the dispersion relations for two theories and found that, in the Maxwell-Chern-Simons theory, a nontrivial Lorentz-breaking modification of the dispersion relations takes place whereas in the self-dual theory, the dispersion relations do not involve Lorentz symmetry breaking, thus, the dual embedding increases Lorentz symmetry breaking.

The natural continuation of this study would contain, first, in more detailed study of the phenomenological applications of the new "mixed" term, second, in its generation via an appropriate coupling of the vector and scalar fields to the spinor matter, similarly to \cite{JK}. 

{\bf{APPENDIX}}
\setcounter{equation}{0}
\renewcommand{\theequation}{A.\arabic{equation}}

In this Appendix, we derive the propagators of the Lorentz-breaking self-dual and Maxwell-Chern-Simons theories.

We start with the Maxwell-Chern-Simons theory whose action (\ref{leff}) is gauge invariant. To obtain the propagator, we add a simplest Feynman-like gauge fixing term $L_{gf}=-\frac{1}{2}(\partial\cdot A)^2$, thus, the Lagrangian takes the form
\begin{equation}
{\cal L}_{eff}^{fixed}=\frac{1}{2}A_{\mu}(\eta^{\mu\nu}\Box+m\epsilon^{\mu\nu\rho}\partial_{\rho})A_{\nu}-\frac{1}{2}\phi(\Box-4v^2)\phi+2\phi
\epsilon^{\mu\nu\rho}v_{\mu}\partial_{\nu}A_{\rho}.
\end{equation}
We can find a propagator in a manner similar to \cite{JTred}. Indeed, the Lagrangian can be presented in the matrix form
\bea
{\cal L}_{eff}^{fixed}=\frac{1}{2}\left(\begin{array}{cc}A^{\mu}\phi\end{array}\right)
\left(\begin{array}{cc}M_{\mu\nu}&T_{\mu}\\
-T_{\nu}&-\Box+M^ 2\\
\end{array}\right)\left(\begin{array}{c}A^{\nu}\\ \phi
\end{array}
\right).\nonumber
\eea 
Here $M^2=4v^2$, the signature is $(-++)$ and 
\bea
\label{oper}
M_{\mu\nu}=\Box\theta_{\mu\nu}-mS_{\mu\nu}+\frac{\Box}{\xi}\omega_{\mu\nu} 
\eea
(after the calculations we put $\xi=1$), and $T_{\nu}=S_{\mu\nu}v^{\mu}$, $S_{\mu\nu}=\epsilon_{\mu\lambda\nu}\partial^{\lambda}$, $\theta_{\mu\nu}=\eta_{\mu\nu}-\omega_{\mu\nu}$, $\omega_{\mu\nu}=\frac{\partial_{\mu}\partial_{\nu}}{\Box}$. The operator determining the theory is
\bea
P=\left(\begin{array}{cc}M_{\mu\nu}&T_{\mu}\\
-T_{\nu}&-\Box+M^2
\end{array}\right).
\eea
The corresponding inverse operator is
\bea
\label{del1}
\Delta=P^{-1}=-\frac{1}{(\Box-M^2)M_{\mu\nu}-T_{\mu}T_{\nu}}
\left(\begin{array}{cc}
-\Box+M^2&T_{\mu}\\
-T_{\nu}&M_{\mu\nu}
\end{array}\right).
\eea
From here we can find the propagators
\bea
\label{props}
<A^{\mu}A^{\nu}>&=&(\Delta_{11})^{\mu\nu}=[(\Box-M^2)M_{\mu\nu}-T_{\mu}T_{\nu}]^{-1}(\Box-M^2)\nonumber\\
<\phi\phi>&=&\Delta_{22}=[(\Box-M^2)M_{\mu\nu}-T_{\mu}T_{\nu}]^{-1}M_{\mu\nu}\nonumber\\
<A^{\mu}\phi>&=&-<\phi A^{\mu}>=\Delta_{12}^{\mu}=-\Delta^{\mu}_{21}=-T_{\nu}[(\Box-M^2)M_{\mu\nu}-T_{\mu}T_{\nu}]^{-1}.
\eea
Thus, all propagators can be expressed in terms of the operator $\Delta=[(\Box-M^2)M_{\mu\nu}-T_{\mu}T_{\nu}]^{-1}$ which we also use to find dispersion relations. Thus, we face a problem to obtain this operator, that is, to solve an equation $P\Delta={\bf 1}$.
To do it, we use an ansatz similar to \cite{JTred}
\bea
\label{ansatz}
\Delta^{\nu\alpha}&=&a_1\theta^{\nu\alpha}+a_2\omega^{\nu\alpha}+a_3S^{\nu\alpha}+a_4\Lambda^{\nu\alpha}+a_5T^{\nu}T^{\alpha}+a_6Q^{\nu\alpha}+a_7Q^{\alpha\nu}+a_8\Sigma^{\nu\alpha}+\nonumber\\&+&a_9\Sigma^{\alpha\nu}+a_{10}\Phi^{\nu\alpha}+a_{11}\Phi^{\alpha\nu},
\eea
where $Q_{\mu\nu}=v_{\mu}T_{\nu}$, $\Lambda_{\mu\nu}=v_{\mu}v_{\nu}$, $\Sigma_{\mu\nu}=v_{\mu}\partial_{\nu}$, $\Phi_{\mu\nu}=T_{\mu}\partial_{\nu}$, $\lambda=v^{\mu}\partial_{\mu}$.

After straightforward but quite tedious calculations we find:
\bea
\label{pmcs}
a_1&=&\frac{1}{(\Box-M^2)(\Box-m^2)};\quad\, a_2=\frac{1}{\Box(\Box-M^2)}-\frac{m^2\lambda^2}{\Box(\Box-m^2)(\Box-M^2){\cal R}};\nonumber\\
a_3&=&\frac{m}{\Box(\Box-m^2)(\Box-M^2)};\quad\, a_4=\frac{m^2}{\Box(\Box-m^2)(\Box-M^2){\cal R}};\nonumber\\
a_5&=&\frac{1}{(\Box-m^2)(\Box-M^2){\cal R}};\quad\, a_6=\frac{m}{(\Box-m^2)(\Box-M^2){\cal R}};\quad\,a_7=-a_6;\\
a_8&=&\frac{m^2\lambda}{\Box(\Box-m^2)(\Box-M^2){\cal R}};\quad\,a_9=-a_8;\quad\,
a_{10}=-\frac{m\lambda}{\Box(\Box-m^2)(\Box-M^2){\cal R}}; \nonumber\\
a_{11}&=&-a_{10}.\nonumber
\eea
Here ${\cal R}=(\Box-M^2)(\Box-m^2)-T^2$. One can verify that for $M^2=0$, the result of \cite{JTred}, where the detailed study of the unitarity, causality and splitting of degrees of freedom into physical and nonplysical ones in the theory governed by this propagator (but with $M^2=0$) is carried out, is reproduced.

Applying the similar method to the self-dual theory with the action (\ref{acao}), we find that the operator determining the quadratic action of the theory is given by the expression (\ref{oper}), with $M_{\mu\nu}$ and $P_{\mu}$ are 
\bea
\label{mn}
M_{\mu\nu}=mS_{\mu\nu}-m^2(\theta_{\mu\nu}+\omega_{\mu\nu}),
\eea
and $T_{\mu}=2mv_{\mu}$ and $M^2=0$. In this case the propagators are given by (\ref{props}), with $M^2=0$ as we had already noted and $M_{\mu\nu}$ is given by (\ref{mn}). The key role is played by the operator $\Delta$ whose expansion again has the form (\ref{ansatz}). Solving again the system for the
coefficients $a_i$ we find
\bea
\label{psd}
a_1&=&\frac{1}{\Box(\Box-m^2)};\quad\, a_2=-\frac{1}{m^2\Box}-\frac{\lambda}{m^2(\Box-m^2)\tilde{\cal R}};\nonumber\\
a_3&=&\frac{1}{m\Box(\Box-m^2)};\quad\, a_4=\frac{1}{\Box(\Box-m^2)\tilde{\cal R}};\nonumber\\
a_5&=&\frac{1}{\Box(\Box-m^2)\tilde{\cal R}};\quad\, a_6=-\frac{1}{m(\Box-m^2)\tilde{\cal R}};\quad\,a_7=-\frac{1}{m(\Box-m^2)\tilde{\cal R}};\nonumber\\
a_8&=&\frac{1}{\Box(\Box-m^2)\tilde{\cal R}};\quad\,a_9=\frac{\lambda}{m^2(\Box-m^2)\tilde{\cal R}};\quad\,
a_{10}=-\frac{1}{m(\Box-m^2)\tilde{\cal R}};\nonumber\\
a_{11}&=&\frac{\lambda}{m\Box(\Box-m^2)\tilde{\cal R}},
\eea
where $\tilde{\cal R}=\Box^2-\Box m^2-T^2$, which is similar to the case of the MCS theory, but with other $T^{\mu}$ (applying the definition of the propagators (\ref{props}) for $M=0$, we find that the $a_2$ contribution will generate a contact term which is known to present always in self-dual theories \cite{embed}).

{\bf Acknowledgements.} This work was partially supported by Conselho Nacional de Desenvolvimento Cient\'{\i}fico e Tecnol\'{o}gico (CNPq). The work by A. Yu. P. has been supported by CNPq-FAPESQ DCR program, CNPq project No. 350400/2005-9.

\end{document}